\documentclass[smus]{snow2e}
\usepackage{epsf}

\begin{document}

\title{Discovery Mass Reach for Topgluons Decaying to $b\bar{b}$ at the Tevatron}

\author{Robert M. Harris\\ {\it Fermilab, Batavia, IL  60510}}

\maketitle

\thispagestyle{empty}\pagestyle{empty}

\begin{abstract} 
%
In topcolor assisted technicolor, topgluons
are massive gluons which couple mainly to top and bottom quarks.
We estimate the mass reach for topgluons decaying to $b\bar{b}$ at the
Tevatron as a function of integrated luminosity.
The mass reach for topgluons decreases with increasing topgluon width, and is 
$0.77 - 0.95$ TeV for Run II (2 fb$^{-1}$) and $1.0-1.2$ TeV for TeV33 (30 
fb$^{-1}$).

\end{abstract}
\section{Topcolor and Topgluons}
Topcolor assisted technicolor~\cite{ref_topcolor} is a model of dynamical
electroweak symmetry breaking in which the top quark is heavy because of a
new dynamics. Topcolor replaces the $SU(3)_C$ of QCD with $SU(3)_1$ for 
the third quark generation and $SU(3)_2$ for the first two generations. 
The additional SU(3) symmetry produces a $<t\bar{t}>$ condensate 
which makes the top quark heavy, and gives rise to a color octet gauge
boson, the topgluon B. The topgluon is expected to be wide 
($\Gamma/M \approx 0.3 - 0.7$) and 
massive ($M \sim 0.5 - 2$ TeV). In hadron collisions it is produced through 
a small coupling to the first two generations, and then decays via a much larger
coupling to the third generation: $q\bar{q} \rightarrow B \rightarrow 
b\bar{b}, t\bar{t}$. Here we estimate the mass reach for topgluons decaying to 
$b\bar{b}$ at the Tevatron.


\section{Signal}

\begin{figure}[tbh]
\hspace*{-0.25in}
\epsffile{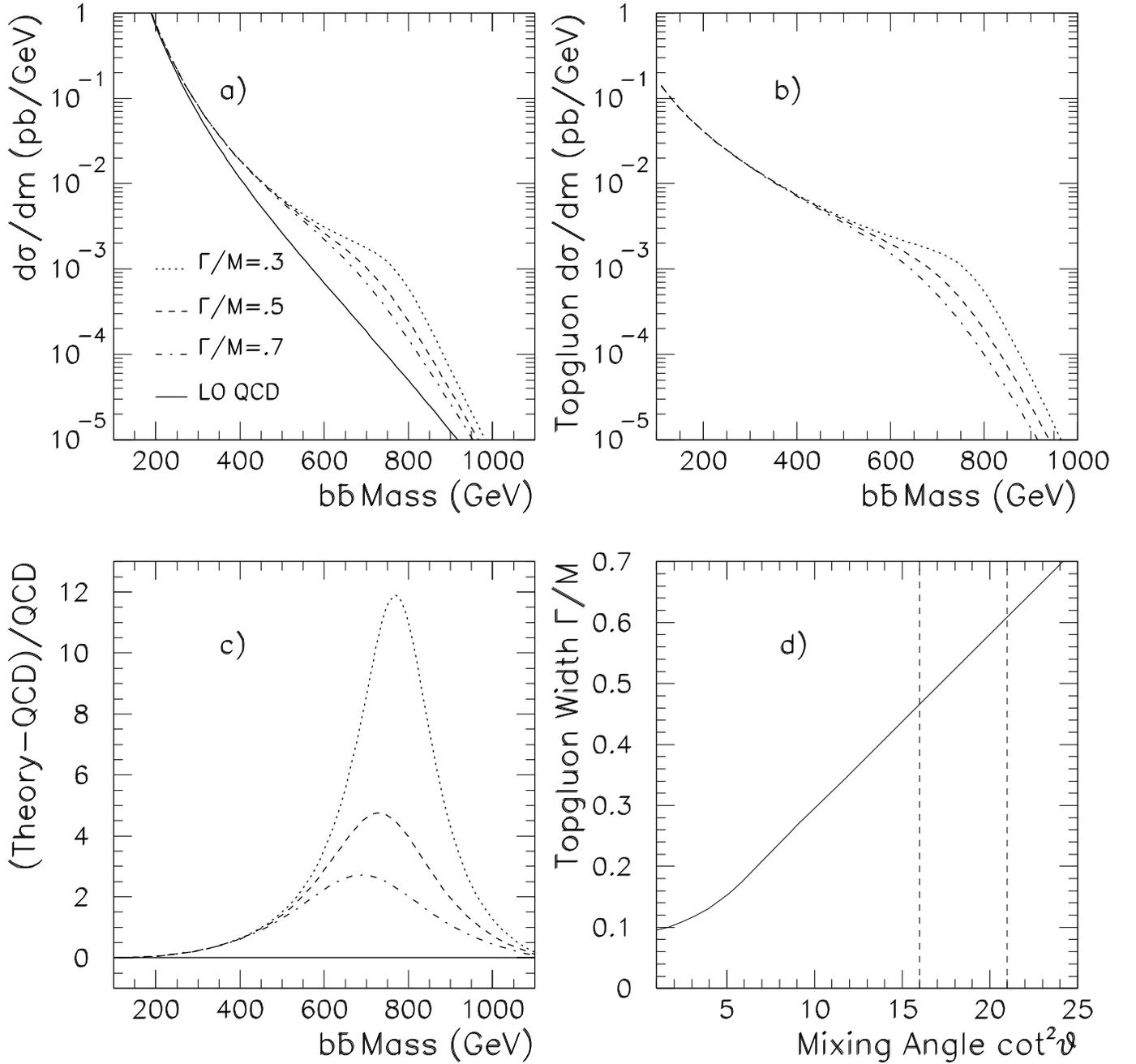}
\caption[]{ Lowest order parton level predictions for an 800 GeV topgluon 
decaying to $b\bar{b}$ displayed as a function of $b\bar{b}$ mass. a) The 
cross section for the LO $b\bar{b}$ background 
from QCD (solid) is compared to the coherent sum of LO QCD and a topgluon
of fractional width $\Gamma/M=0.3$ (dots), 0.5 (dashes) and 0.7 (dotdash).
In b) the QCD prediction has been subtracted leaving only the topgluon signal
and the interference between QCD and topgluons (constructive beneath peak, 
destructive above peak). c) The fractional deviation above the QCD
prediction produced by the presence of a topgluon. d) The solid curves give 
the topgluon width as a function of the mixing angle between $SU(3)_1$ and 
$SU(3)_2$ for an 800 GeV topgluon. The vertical dashed lines indicate the 
theoretically preferred range of mixing angle~\cite{ref_topg_range}.}
\label{fig_parton}
\end{figure}

The sub-process cross section for $q\bar{q} \rightarrow b\bar{b}$ from both QCD
and topgluons  is given by~\cite{ref_lane}
\begin{equation}
\frac{d\hat{\sigma}}{d\hat{t}} = \frac{2\pi\alpha_s^2}{9\hat{s}^2} (1
-\cos^2\theta^*)\left| 1 - \frac{\hat{s}}{\hat{s} - M^2 +
i\sqrt{\hat{s}}\Gamma} \right|^2
\label{eq_xsec}
\end{equation}
for a topgluon of mass $M$ and width $\Gamma$ given by 
\begin{equation}
\Gamma = \frac{\alpha_s M}{6} \left[ 4 \tan^2\theta + \cot^2\theta \left( 1 
+ \beta_t \left( 1-\frac{m_t^2}{M^2} \right) \right) \right]
\label{eq_width}
\end{equation}
where $\alpha_s$ is the strong coupling, $\hat{s}$ and $\hat{t}$ are subprocess
Mandelstam variables, 
$\theta$ is the mixing angle between $SU(3)_1$ and $SU(3)_2$, 
$\theta^*$ is the scattering angle between the bottom quark and the initial state quark in the
center of mass frame, 
$\beta_t=\sqrt{1-4m_t^2/M^2}$, and $m_t$ is the top quark mass.
Topcolor requires $\cot^2\theta>>1$ to make the top quark heavy.
In Eq.~\ref{eq_width}, the first term in square brackets is for four light
quarks, and the second term has two components, the first for the bottom
quark and the second for massive top quarks.
In Eq.~\ref{eq_xsec}, the 1 inside the absolute
value brackets is for the normal QCD process $q\bar{q} \rightarrow g \rightarrow
b\bar{b}$.  The other term inside the brackets is the Breit-Wigner topgluon
resonance term for the process $q\bar{q} \rightarrow B \rightarrow b\bar{b}$.
The two processes interfere constructively to the left of the mass peak and
destructively to the right of the mass peak. CDF has done a preliminary 
search~\cite{ref_pbarp} in which the interference between normal gluons and 
topgluons was modeled in the opposite way: destructive to the left of the 
mass peak and constructive to the right. That model of the interference, 
hybrid model C in reference~\cite{ref_hybridc}, is inappropriate for topcolor 
assisted technicolor and is replaced by Eq.~\ref{eq_xsec}.
In Fig.~\ref{fig_parton} we have convoluted Eq.~\ref{eq_xsec} with CTEQ2L
parton distributions~\cite{ref_cteq} to calculate 
the QCD background and topgluon signal
for the case of an 800 GeV topgluon in $p\bar{p}$ collisions at $\sqrt{s}=1.8$ 
TeV. Fig.~\ref{fig_parton} also includes 
the QCD process $gg\rightarrow b\bar{b}$ which is only significant at low mass.
In Fig.~\ref{fig_parton}1a we plot the differential cross section $d\sigma/dm$,
where $m$ is the invariant mass of the $b\bar{b}$ system.
We require both $b$ quarks to have
pseudorapidity $|\eta|<2$ and the $b\bar{b}$ system to have center of 
mass scattering angle $|\cos\theta^*|<2/3$.
A clear distortion of the QCD $b\bar{b}$ 
spectrum is caused by the presence of a topgluon in Fig.~\ref{fig_parton}a. 
After subtraction of the QCD
background, Fig~\ref{fig_parton}b shows that the signal has a very long high 
tail to low masses, 
caused by the combination of constructive interference and parton
distributions that rise rapidly as the $b\bar{b}$ mass decreases.  The
tail is significantly larger than the peak, as seen in Fig.~\ref{fig_parton}b.
Nevertheless, the ratio between the topgluon signal and the QCD background,
displayed in Fig.~\ref{fig_parton}c, displays a noticeable peak close to the
topgluon mass. 

In Fig.~\ref{fig_pythia_qfl} we have repeated the calculations of
Fig.~\ref{fig_parton} using the PYTHIA Monte Carlo~\cite{ref_pythia}, including QCD radiation,
and a simulation of the CDF detector. The mass peaks due to a topgluon are 
still visible when compared to QCD on a linear scale in
Fig.~\ref{fig_pythia_qfl}c.  Similar calculations have been performed
for the masses 400, 600, and 1000 GeV.

\section{Backgrounds and B-Tagging}
The QCD background we have considered so far is only the lowest order 
processes $q\bar{q} \rightarrow b\bar{b}$ and $gg \rightarrow b\bar{b}$.  
For an analysis in which we only tag one of the two b quarks,
the background should also includes contributions from final state gluon 
splitting to $b\bar{b}$, flavor excitation where an initial state gluon
splits into $b\bar{b}$ and one of the two b quarks undergoes a hard scatter,
and contributions from jets faking a b-tag and from charm. To get the most 
realistic estimate of the background we use CDF run 1A data in which we require
at least one of the two leading jets to be tagged as a bottom quark. The b-tag
requires a displaced vertex in the secondary vertex detector~\cite{ref_b_tag}.
The $b\bar{b}$ reconstruction efficiency when we required at least
a single b-tag 
\onecolumn

\begin{figure}[tbh]
\hspace*{-0.25in}
\epsffile{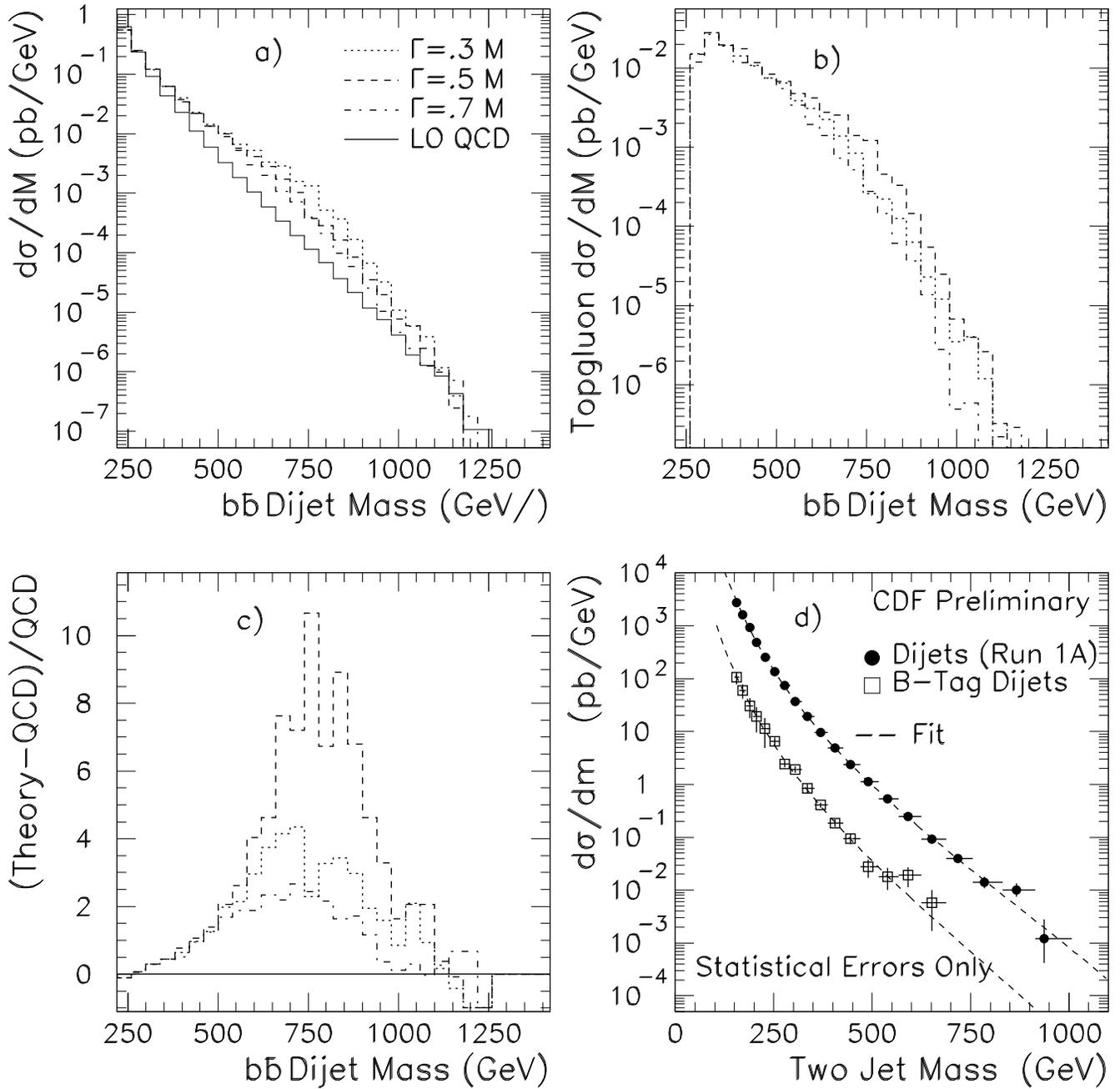}
\caption[]{ Simulation of an 800 GeV topgluon and measurement of background
in the CDF detector.  a), b) and c) are the same as in
Fig.~\ref{fig_parton} except they include QCD radiation from PYTHIA and a 
CDF detector simulation. d) Dijet mass data (solid points) and b-tagged dijet 
mass data (open boxes) both from run 1A only, are compared to a 
parameterized fit to the data (curve).}
\label{fig_pythia_qfl}
\end{figure}

\begin{figure}[tbh]
\hspace*{-0.25in}
\epsffile{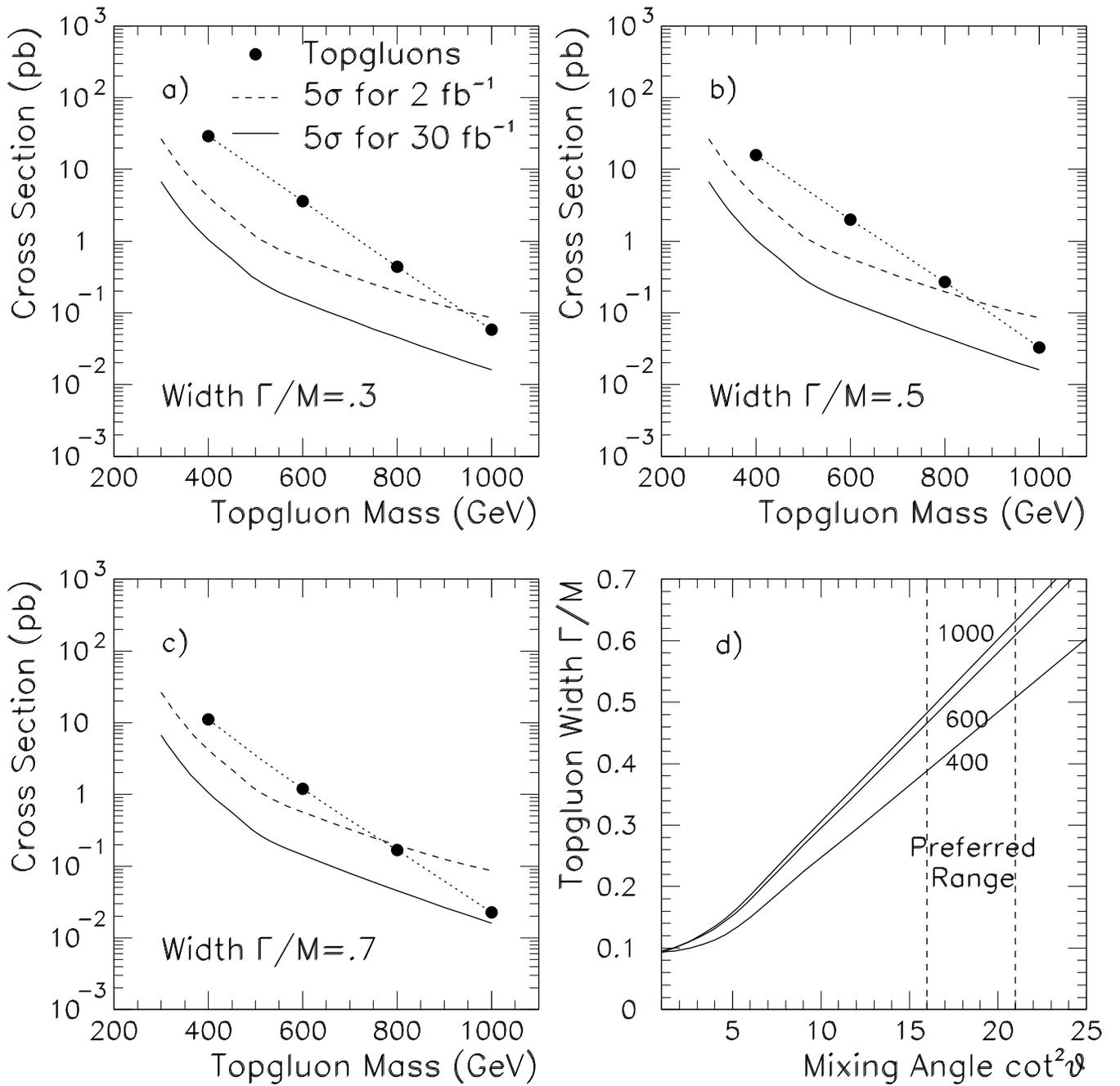}
\caption[]{ The mass reach for $b\bar{b}$ decays of topgluons of width a)
0.3 M, b) 0.5 M, and c) 0.7 M.  The predicted cross section for
topgluons (points) is compared to the 5$\sigma$ discovery reach of the Tevatron
with a luminosity of 2 fb$^{-1}$ (dashed) and 30 fb$^{-1}$ (solid). All
cross sections are for $b\bar{b}$ with $|\eta|<2$, $|\cos\theta^*|<2/3$, and
invariant mass within 25\% of the topgluon peak.  d) The solid curves give 
the topgluon width as a function of the mixing angle between $SU(3)_1$ and 
$SU(3)_2$ for 3 different topgluon masses. The vertical dashed lines indicate 
the theoretically preferred range of mixing angle~\cite{ref_topg_range}.}
\label{fig_reach}
\end{figure}

\twocolumn
\noindent
was $0.25$ in run 1A~\cite{ref_pbarp}. 
In Fig.~\ref{fig_pythia_qfl}d we show CDF run 1A data on both the untagged 
dijet mass spectrum and
the b-tagged dijet mass spectrum compared to a parameterized
fit~\cite{ref_pbarp}. Notice that the single b-tagged 
dijet mass spectrum is 
about an order of magnitude higher than the simulated background from direct
$b\bar{b}$ in 
Fig.~\ref{fig_pythia_qfl}a. 
We estimate that the single b-tagged data is 
roughly 20\% fakes, 30\% charm, 50\% bottom, and only 1/5 of the bottom
component is direct $b\bar{b}$ and the rest is gluon splitting and flavor
excitation~\cite{ref_pbarp}. We fit the b-tagged dijet mass spectrum to the 
functional form, $d\sigma/dm = A(1-m/\sqrt{s})^N/m^p$, with parameters $A$, 
$N$ and $p$. The fit is used to estimate the background to topgluons 
when calculating the discovery reach.  
\section{Discovery Mass Reach}

To calculate the discovery mass reach we integrate both the fully simulated 
topgluon signal 
cross section and the b-tagged dijet background parameterization within the
range $0.75M<m<1.25M$.
The resulting total topgluon signal in the $b\bar{b}$ channel is shown
in Fig.~\ref{fig_reach}. The resulting background rate
in this mass range is used to find the 5 $\sigma$ discovery cross section. This
is conservatively defined as the cross section which is
above the background by 5 $\sigma$, where $\sigma$ is the statistical error on 
the measured cross section (not the background).  For example, if the 
background were zero events the $5\sigma$ discovery rate would be 25 events.  
To obtain the discovery cross section we used both the luminosity
and the 25\% $b\bar{b}$ reconstruction efficiency. 
In Fig.~\ref{fig_reach} we compare the
topgluon signal cross section to the 5 $\sigma$ discovery cross section for two 
different luminosities: 2 fb$^{-1}$ for Tevatron collider run II and 
30 fb$^{-1}$ for TeV33. The topgluon discovery mass reach, defined as the mass 
at which a topgluon would be discovered with a 5$\sigma$ signal, 
is tabulated in Table I as a function of integrated luminosity and topgluon 
width.
The mass reach decreases with increasing width, caused by worsening 
signal to background within the search window. The width as a function
of mixing angle, from Eq.~\ref{eq_width}, is shown in Figs.~\ref{fig_parton}d 
and \ref{fig_reach}d. Also shown is the preferred theoretical range
for the mixing angle $\cot^2\theta$, determined from the topcolor model and 
constraints from other data~\cite{ref_topg_range}, which then implies
an allowed width of the topgluon somewhere in the range 
$\Gamma/M \approx 0.3 - 0.7$.

\section{Systematics and Improvements}
In this analysis, we have not included any systematic uncertainties on the
measured signal, and we have assumed
that the shape and magnitude of the single b-tagged background spectrum will
be well understood.  Also, we have approximated the 
background by an extrapolation of existing b-tagged data into a higher mass 
region. Adding systematics on the signal and the background will 
decrease the mass reach of a real search. 
However, we anticipate a factor of two improvement in
b-tagging efficiency when the run 1B algorithm is used, and further 
improvements are possible in Run II and TeV33. Also, this analysis only 
considers single b-tagging; an 
analysis which tags both $b$ quarks from the topgluon decay will likely have a 
significantly better discovery reach because the backgrounds from gluon 
splitting, flavor excitation, and fakes will be significantly reduced.  
In order to have a background estimate from the data, we did the analysis 
for a center of mass energy of $\sqrt{s}=1.8$ TeV, while Run II and TeV33 will
be at $\sqrt{s}=2.0$ TeV which could provide a 10\% larger mass reach.
We have also not done a maximum  likelihood fit, which would be a more 
sensitive statistical test for the presence of a signal.  Adding these 
improvements should 
increase the mass reach of a real search. The positive consequences of future
improvements will likely exceed the negative consequences of 
neglecting systematics. We believe our analysis gives a conservative estimate 
of the mass reach.
\\
\\
\\
Table I: The 5$\sigma$ discovery mass reach of the Tevatron in Run II
(2 fb$^{-1}$) and TeV33 (30 fb$^{-1}$) for a toplguon decaying to $b\bar{b}$ 
as a function of its fractional width ($\Gamma/M$).
\begin{table}[tbh]
\begin{center}
\begin{tabular}{|c|c|c|}\hline 
Width & \multicolumn{2}{c|}{Mass Reach} \\
$\Gamma/M$ & 2 fb$^{-1}$ & 30 fb$^{-1}$ \\ \hline
0.3          & 0.95 TeV &  1.2 TeV \\
0.5          & 0.86 TeV &  1.1 TeV \\
0.7          & 0.77 TeV &  1.0 TeV \\ \hline
\end{tabular}
\end{center}
\end{table}

\section{Summary and Conclusions} 
We have used a full simulation of topgluon production and decay to $b\bar{b}$,
and an extrapolation of the b-tagged dijet mass data~\cite{ref_pbarp}, to
estimate the topgluon discovery mass reach in a $b\bar{b}$ resonance search. 
The topgluon discovery mass reach, 
$0.77-0.95$ TeV for Run II and $1.0-1.2$ TeV for TeV33, covers a significant 
part of the expected mass range ($\sim0.5 - 2$ TeV).  
For comparison, the mass reach in the $t\bar{t}$ channel is estimated to be
$1.0-1.1$ TeV for Run II and $1.3 - 1.4$ TeV for TeV33~\cite{ref_ttbar}.
This is greater than the mass reach in the $b\bar{b}$ channel primarily 
because of smaller $t\bar{t}$ backgrounds.
If topgluons exist, there is a good chance we will find
them at the Tevatron, beginning the investigation into the origins of 
electroweak symmetry breaking.

\end{document}